\documentclass[showpacs,nofootinbib,amsmath,amssymb,superscriptaddress]{revtex4}

\usepackage{pstricks}
\usepackage{graphicx}
\usepackage{epsfig}
\usepackage{hhline,braket}
\usepackage{amsmath,amssymb,amsfonts}
\newcommand{\lms}{\Lambda_{\overline{\mbox{\tiny{MS}}}}}

\usepackage{amsthm}
\usepackage{float,braket}

\newcommand{\MSbar}{\overline{\mbox{MS}}}

\newcommand{\p}{\partial}

\begin{document}
\title{{\bf\Large Indirect lattice evidence for the Refined Gribov-Zwanziger formalism and the gluon condensate $\braket{A^2}$ in the Landau gauge}}
\author{\bf D. Dudal}
\email{david.dudal@ugent.be}
\affiliation{Ghent University, Department of Physics and Astronomy, Krijgslaan 281-S9, B-9000 Gent, Belgium}
\author{\bf O. Oliveira }
\email{orlando@teor.fis.uc.pt}
\affiliation{Departamento de F\'{\i}sica, Universidade de Coimbra, P-3004-516 Coimbra, Portugal}
\author{\bf N. Vandersickel }
\email{nele.vandersickel@ugent.be}
\affiliation{Ghent University, Department of Physics and Astronomy, Krijgslaan 281-S9, B-9000 Gent, Belgium}
%\date{\today}
%
\begin{abstract}
We consider the gluon propagator $D(p^2)$ at various lattice sizes and spacings in the case of pure SU(3) Yang-Mills gauge theories using the Landau gauge fixing. We discuss a class of fits in the infrared region in order to (in)validate the tree level analytical prediction in terms of the (Refined) Gribov-Zwanziger framework. It turns out that an important role is played by the presence of the widely studied dimension two gluon condensate $\braket{A^2}$. Including this effect allows to obtain an acceptable fit up to 1 \'{a} 1.5 GeV, while corroborating the Refined Gribov-Zwanziger prediction for the gluon propagator.  We also discuss the infinite volume extrapolation, leading to the estimate $D(0)=8.3\pm0.5~\text{GeV}^{-2}$. As a byproduct, we can also provide the prediction $\braket{g^2 A^2}\approx 3~\text{GeV}^2$ obtained at the renormalization scale $\mu=10~\text{GeV}$.

\end{abstract}
\maketitle

\section{Introduction}
Although confinement of gluons in pure Yang-Mills gauge theories should be understood in a gauge invariant setting, one should also observe in some way the messengers of confinement in gauge variant quantities. In particular, let us assume that we have fixed our gauge freedom by means of the Landau gauge, $\p_\mu A_\mu^a=0$. We expect that the nonperturbative physics responsible for confinement will also reveal their influence on the $n$-point correlation functions of the gauge fixed theory. The most elementary, albeit already nontrivial, example of such correlation function is the gluon propagator. Due to the transverse nature of the Landau gauge, we can write
\begin{equation}\label{I1}
    \braket{ A_\mu^a(p) A_\nu^b(-p)}= D(p^2)\delta^{ab}\left(\delta_{\mu\nu}-\frac{p_\mu p_\nu}{p^2}\right)\,,
\end{equation}
and focus attention on the scalar quantity $D(p^2)$. This quantity has been the topic of a lot of investigations in the recent past, be it from numerical \cite{
Leinweber:1998im,Bonnet:2001uh,Furui:2004cx,Silva:2004bv,Cucchieri:2007zm,Oliveira:2008uf,Oliveira:2009nn,Bornyakov:2009ug,Cucchieri:2007md,
Cucchieri:2007rg,Cucchieri:2009zt,Maas:2008ri,Maas:2009ph,Bogolubsky:2007ud,Bogolubsky:2009dc} or analytical viewpoint \cite{Gribov:1977wm,Zwanziger:1989mf,Zwanziger:1992qr,Zwanziger:2001kw,Zwanziger:2003cf,Gracey:2006dr,Dudal:2005na,Dudal:2007cw,Dudal:2008sp,Sorella:2009vt,Alkofer:2000wg,Alkofer:2003jj,
Lerche:2002ep,Pawlowski:2003hq,Huber:2009tx,Kondo:2002cx,Aguilar:2004sw,Aguilar:2008xm,Binosi:2009qm,Fischer:2008uz,Schleifenbaum:2006bq,Pene:2009iq}. All data and analytical estimates agree on the fact that the gluon is infrared suppressed. There is still some discussion whether it actually vanishes at zero momentum or not, but most lattice data seems to indicate that it does not. \\

An important asset in the computation $D(p^2)$ is the role played by Gribov copies \cite{Gribov:1977wm,Singer:1978dk}. In principle, the gauge configuration $A_\mu$ is in the (absolute) Landau gauge if it corresponds to the absolute minimum of the functional
\begin{equation}\label{I2}
    R[A]\equiv A^2_{\min}=\min_{u\in \text{SU(N)}} \int d^dx (A_\mu^u)^2\,.
\end{equation}
The set of absolute minima defines $\Lambda$, the so-called Fundamental Modular Region (FMR). It is then an easy exercise to show that $A_\mu$ is part of the Gribov region, defined as
\begin{align}
\Omega = \{ \; A; \; \partial_{\mu} A^a_{\mu}=0, \; {\cal M}^{ab} > 0 \; \}   \,,  \label{om}
\end{align}
with ${\cal M}^{ab}$ the (Hermitian) Faddeev-Popov operator, defined by
\begin{equation}
{\cal M}^{ab}=-\partial_\mu D_\mu^{ab}  = -(\delta^{ab}\partial^2+g\,f^{abc} A_\mu^c \partial_\mu)\,.
\end{equation}
Notice that the requirement ${\cal M}^{ab} > 0$ is necessary to remove many redundant gauge configurations, as the transversality condition $\p_\mu A_\mu=0$ has multiple solutions along each gauge orbit. Said otherwise, the Landau gauge is plagued by the existence of Gribov copies. It is important to mention that $A_\mu \in \Omega$ does not necessarily mean that $A_\mu$ corresponds to the absolute minimum of $R[A]$; it can also constitute a relative minimum. Said otherwise, the FMR $\Lambda$ is a subset of the Gribov region $\Omega$. This means that the Gribov region $\Omega$ still contains gauge copies, see for instance \cite{semenov,vanBaal:1991zw,Dell'Antonio:1991xt}. It can be shown that $\Omega$ is convex, bounded in all directions and that is crossed by any gauge orbit \cite{Zwanziger:2003cf,Dell'Antonio:1989jn,Dell'Antonio:1991xt}.\\

In lattice computations, one fixes the Landau gauge numerically by searching for the ``best'' solution of the minimization procedure for $R[A]$. As such, one hopes to bring each configuration as close as possible to a gauge equivalent one in $\Lambda$.\\

In the continuum, it appears to be an incredibly difficult task to implement the absolute Landau gauge. In a first approximation, one uses the perturbative Faddeev-Popov action,
\begin{eqnarray}
S_{YM+gf} &=& \frac{1}{4}\int d^d x F^a_{\mu\nu} F^a_{\mu\nu}+\int d^d x\,\left( b^a \p_\mu A_\mu^a +\overline c^a \p_\mu D_\mu^{ab} c^b \right)\,,
\end{eqnarray}
which just implements $\p_\mu A_\mu=0$, by means of the equation of motion of the auxiliary $b$-field. The corresponding Jacobian determinant is represented by the ghost term $\overline c^a \p_\mu D_\mu^{ab} c^b$. This approach completely ignores the existence of Gribov copies, but it is perfectly well-suited for the perturbative quantization of gauge theories.\\

A second approximation involves the restriction of the allowed gauge configurations to the Gribov region $\Omega$, which already partially resolves the issue of gauge copies. It was worked out at lowest order in a saddle point approximation in \cite{Gribov:1977wm} and later on generalized to all orders in \cite{Zwanziger:1989mf,Zwanziger:1992qr}. We shall skip the details, and just mention the result, i.e.~the local action
\begin{eqnarray}\label{GZ}
S_{GZ}&=&S_{YM+gf}\\&+& \int d^d x \left( \overline \varphi_\nu^{ac} \p_\mu \left( D_\mu^{ab} \varphi_\nu^{bc} \right) - \overline \omega_\nu^{ac} \p_\mu \left( D_\mu^{ab} \omega_\nu^{bc} \right) - g f^{abc} \p_\mu \overline \omega_\nu^{ak}    D_\mu^{bd} c^d  \varphi_\nu^{ck}  - \gamma ^{2}g f^{abc}A_{\mu }^{a}\left(\varphi _{\mu }^{bc}+\overline{\varphi }_{\mu }^{bc}\right) -d\left(N^{2}-1\right) \gamma^{4} \right) \,,\nonumber
\end{eqnarray}
which contains the (Gribov) mass parameter $\gamma^2$. This parameter is not free, but self-consistently fixed by means of the so-called horizon condition \cite{Zwanziger:1989mf,Zwanziger:1992qr}, which reads in its local version $\braket{gf^{abc}A_\mu^a(\varphi_{\mu}^{bc}+\overline\varphi_\mu^{bc})}=-2d(N^2-1)\gamma^2$. Upon solving, the horizon condition shall give $\gamma^2\propto \Lambda_{QCD}$. A crucial feature of this local formulation of the restriction is that we can control its ultraviolet behaviour, i.e.~the action defines a renormalizable quantum field theory \cite{Zwanziger:1992qr,Dudal:2005na,Maggiore:1993wq,Dudal:2010fq}. As such, a consistent computational framework is obtained. So far, no one has been able to improve upon this restriction, in particular it is unclear whether it would be possible to implement the restriction to the FMR $\Lambda$ to completely overcome the gauge fixing ambiguity. We can only refer to the conjecture of \cite{Zwanziger:2003cf} stating that at the level of expectation values, no difference will be found upon restricting to $\Omega$ or to its subset $\Lambda$. Anyhow, the presence of the mass parameter $\gamma^2\propto \Lambda_{QCD}$ will clearly be generating nonperturbative effects in e.~g.~gluon and ghost propagator.\\

A well-known important source of nonperturbative effects in gauge theories are condensates, viz.~vacuum expectation values of certain local operators. Next to the famous gauge invariant condensate $\braket{F_{\mu\nu}^2}$, of paramount importance for phenomenological applications \cite{Shifman:1978bx}, recent years\footnote{The $d=2$ gluon condensate was already considered in \cite{Greensite:1985vq,Lavelle:1988eg}.} have also witnessed an increased interest in the dimension two condensate $\braket{A^2}$ in the Landau gauge \cite{Gubarev:2000eu,Gubarev:2000nz}, and related to it the issue of $1/Q^2$ power corrections \cite{Narison:2005hb}. The latter corrections would correspond to an extension of the usual SVZ sum rule study of physical correlators. Some important early contributions to this field of research can be found in, for example, \cite{Grunberg:1997ud,Akhoury:1997ys,Akhoury:1997by,Gubarev:1998ew,Chetyrkin:1998yr,Zakharov:1999jj,Burgio:1997hc,Bali:1999ai,Chernodub:2000bk}. These works were based on renormalon analyses, lattice considerations of the interquark potential and condensates, nonperturbative short distance physics, $\ldots$.  Also at the propagator level such power corrections were identified in \cite{Boucaud:2000ey}. \\

From the definition \eqref{I2}, it is clear that $\braket{A^2_{\min}}$ is a gauge invariant quantity by construction.  This leads very naturally to the introduction of $\braket{A^2}$ in the Landau gauge since we can write \cite{Lavelle:1995ty}
\begin{eqnarray}
A_{\min }^{2}
&=&\frac{1}{2}\int d^{d}x\left[ A_{\mu }^{a}\left( \delta _{\mu \nu }-\frac{\partial _{\mu }\partial _{\nu }}{\partial ^{2}}\right) A_{\nu
}^{a}-gf^{abc}\left( \frac{\partial _{\nu }}{\partial ^{2}}\partial
A^{a}\right) \left( \frac{1}{\partial ^{2}}\partial {A}^{b}\right)
A_{\nu }^{c}\right] \;+O(A^{4})\,,  \label{min1}
\end{eqnarray}
from which it easily follows that $\braket{A^2_{\min}}=\braket{A^2}$ in the Landau gauge. This condensate then made its appearance in a variety of works, see e.g.~ \cite{Dudal:2005na,
Boucaud:2001st,Boucaud:2002nc,Boucaud:2008gn,Verschelde:2001ia,Dudal:2002pq,Dudal:2003vv,Browne:2003uv,Vercauteren:2007gx,Chernodub:2008kf,Dudal:2009tq,Kondo:2001nq,Li:2004te,Gubarev:2005it,Kekez:2005ie,Andreev:2006vy,RuizArriola:2004en,RuizArriola:2006gq,Megias:2009mp}. In the works \cite{Gubarev:2000eu,Gubarev:2000nz}, the relation was explored between this condensate and magnetic degrees of freedom, which are generally believed to play an important role for confinement.  Recently, this was further investigated by looking at the electric and magnetic components of $\braket{A^2}$ at finite temperature, hinting towards an interesting connection with the phase diagram \cite{Chernodub:2008kf}.\\

Measurements of $\braket{A^2}$ at $T=0$ have been obtained using the lattice gluon propagator and the Operator Product Expansion (OPE) in \cite{Boucaud:2008gn}, based on earlier work \cite{Boucaud:2001st,Boucaud:2002nc}, giving the following estimate
\begin{equation}\label{A3}
\braket{g^2A^2} =5.1^{+0.7}_{-1.1}~\text{GeV}^2
\end{equation}
at the renormalization scale $\mu=10~\text{GeV}$. $\braket{A^2}$ also appeared as a source of power corrections in e.g.~\cite{RuizArriola:2006gq,Megias:2009mp}. An independent estimate using the OPE and the quark propagator in a quenched lattice simulation gave \cite{RuizArriola:2004en}
\begin{equation}\label{A3bis}
\braket{g^2A^2} =4.4\pm0.4~\text{GeV}^2\,.
\end{equation}
An ab initio calculation of $\braket{A^2}$ was presented in \cite{Verschelde:2001ia,Browne:2003uv}. It was shown that it is possible to construct an effective potential for $\braket{A^2}$ which is consistent with the renormalization (group) \cite{Verschelde:2001ia,Dudal:2002pq}. A nonvanishing condensate due to dimensional transmutation was favoured as it lowered the vacuum energy. Using a resummation of Feynman diagrams, more evidence for $\braket{A^2}\neq0$ was given in  \cite{Dudal:2003vv}.\\

The extension of the effective potential formalism to the Gribov-Zwanziger case was first tackled in \cite{Dudal:2005na}. More recently, it also became clear that other $d=2$ condensates can play an important role in the Gribov-Zwanziger formalism. When the dynamics of the extra fields is taken into account, next to $\braket{A^2}$ other dimension two condensates appear quite naturally \cite{Dudal:2007cw,Dudal:2008sp}, and these condensates alter the behaviour of the propagators quite drastically. In particular, the $d=2$ condensates related to the auxiliary fields $\left\{\varphi_{\mu}^{ab}, \overline\varphi_{\mu}^{ab}, \omega_{\mu}^{ab}, \overline\omega_{\mu}^{ab}\right\}$ give a ghost propagator behaving like $\sim 1/p^2$ in the infrared, while the gluon propagator is suppressed and tends to a nonzero constant at very low momentum. This framework is now known as the Refined Gribov-Zwanziger (RGZ) formalism, which is a dynamical improvement of the original Gribov-Zwanziger approach. In \cite{Dudal:2007cw,Dudal:2008sp}, the effects of the condensate $\braket{\overline\varphi_{\mu}^{ab}\varphi_{\mu}^{ab}-\overline\omega_{\mu}^{ab}\omega_{\mu}^{ab}}$ were explored by means of variational perturbation theory. $\braket{A^2}$ was left out of this analysis for simplicity, as the qualitative conclusions about the deep infrared behaviour were not depending on this condensate, but it was already discussed in \cite{Dudal:2008sp} that in principle it can be included. In \cite{Dudal:2010}, a more complete treatment will be presented. We shall not dwell upon details here, but focus on the form of the tree level propagator in the presence of these condensates, which is
\begin{equation}
   D(p^2) = \frac{p^2 + M^2}{p^4 + \left(M^2 + m^2\right) p^2 + 2 g^2 N \gamma^4 + M^2 m^2  }\,,
   \label{RZGtreelevel}
\end{equation}
where $M^2$ is the mass scale related to the $d=2$ condensates in $\left\{\varphi_{\mu}^{ab}, \overline\varphi_{\mu}^{ab}, \omega_{\mu}^{ab}, \overline\omega_{\mu}^{ab}\right\}$ , $m^2$ to $\braket{A^2}$ and $\gamma^4$ is the Gribov parameter. We shall introduce the shorthand $\lambda^4=2 g^2 N \gamma^4 + M^2 m^2$. \\

The aim of this paper is to find out whether the propagator \eqref{RZGtreelevel} can reproduce not only qualitatively the gluon propagator, but that it also works out well at the \emph{quantitative} level. We shall therefore analyze the lattice gluon propagator in pure SU(3) Yang-Mills gauge theories and investigate to what extent the propagator \eqref{RZGtreelevel} can match the data, by treating the mass scales $m^2$, $M^2$ and $\gamma^4$ as fitting parameters. \\

The paper is organized as follows. In Section II, we summarize the introduction of the lattice gluon propagator and discuss its renormalization. In Section III, we analyze the propagator and discuss a class of fits related to the RGZ propagator \eqref{RZGtreelevel}. It shall turn out that none of the parameters $m^2$, $M^2$ or $\gamma^4$ can be put equal to zero to find a decent fit, which thereby shows that the analytical (Refined) Gribov-Zwanziger restriction is in compliance with the lattice data, but that it needs to be complemented with the condensate $\braket{A^2}$, as well as the RGZ mass scale $M^2$, related to a new $d=2$ condensate in the $\left\{\varphi_{\mu}^{ab}, \overline\varphi_{\mu}^{ab}, \omega_{\mu}^{ab}, \overline\omega_{\mu}^{ab}\right\}$ fields \cite{Dudal:2008sp}, which is crucial to find $D(0)>0$. We shall also derive an estimated value for the condensate $\braket{g^2A^2}$, and we shall see that it compares acceptably well with the OPE estimates \eqref{A3} and \eqref{A3bis}. In addition, we can also derive an infrared gluon mass scale, which lies in the same ballpark as other values in the literature. We end with conclusions in Section IV.

\section{The Lattice gluon propagator and renormalization procedure}
Lattice QCD simulations are performed on a finite 4D torus. Therefore, either the infinite volume limit should
be taken or the simulations must be performed in a sufficiently large volume. Of course, the precise meaning of a sufficiently
large volume depends on the problem to be addressed. Considering pure Yang-Mills theory and taking the mass
of the lightest glueball, i.e.  $M_{glueball} \sim 1.7$ GeV \cite{Chen:2005mg,Mathieu:2008me} as a typical hadronic scale, the
corresponding length scale is $L \sim 0.1$ fm. However, the scale at which nonperturbative physics sets in is already at
$\sim 1$ fm. To investigate nonperturbative physics and, in particular, the infrared gluon propagator,
one should consider volumes well above the 1 fm scale.\\

In this work we will analyze the propagator computed from the lattices described in TABLE \ref{LatticeSetup}.
Of the three $\beta$ values, $\beta = 6.0$ will be used to perform an extrapolation to the infinite volume
limit, whilst the Berlin-Moscow-Adelaide at $\beta = 5.7$ and $\beta = 6.2$ will be used to cross-check the
final results. In what concerns the computation of the gluon propagator, we will use standard definitions which
can be found in, for example, \cite{Leinweber:1998im,Silva:2004bv}, and as such, it will not be repeated here. The gauge configurations were generated using version 6 of the MILC \cite{MILC}.\\

The lattice data for the propagators computed at $\beta = 6.0$ and $\beta = 6.2$ were chosen as follows.
For momenta higher than $\sim 1$ GeV,  only those momenta which survive the conic cut \cite{Leinweber:1998im} are used. In this
way we avoid the problems associated with the breaking of rotational invariance. For momenta below $\sim 1$
GeV, all momenta were included in the analysis. In this way, we hope to have obtained a decent description in the infrared.\\

\begin{table}[t]
   \begin{center}
   \begin{tabular}{l@{\hspace{0.5cm}}c@{\hspace{0.3cm}}c@{\hspace{0.3cm}}c@{\hspace{0.3cm}}c@{\hspace{0.3cm}}c}
      \toprule
       \multicolumn{6}{c}{$\beta = 5.7$ \hspace{2cm} $a = 0.1838$ fm} \\
       \hline
      $L$                 &    64    &  72    &  80    &  88    &  96 \\
      $aL$ (fm)   &  11.8  & 13.2 &  14.7 & 16.2 & 17.6 \\
      \# Conf       &   14     & 20    &  25    &  68   &  67   \\
      \toprule
       \multicolumn{6}{c}{$\beta = 6.0$ \hspace{2cm} $a = 0.1016$ fm} \\
       \hline
      $L$                 &    32    &  48    &  64    &  80    &   \\
      $aL$ (fm)   &  3.25  & 4.88 &  6.50 & 8.13 &  \\
      \# Conf       &   126   & 104  &  120  &  47   &    \\
      \toprule
       \multicolumn{6}{c}{$\beta = 6.2$ \hspace{2cm} $a = 0.07261$ fm} \\
       \hline
      $L$                 &   48    &  64    &   &      &   \\
      $aL$ (fm)   &  3.49 & 4.65 &    &     &  \\
      \# Conf       &   88   & 99     &    &    &    \\
      \toprule
   \end{tabular}
   \caption{The lattice setup. For the conversion to physical units we took the lattice spacing
                  measure from the string tension \cite{Bali:1992ru}. The first set of configurations, i.e.~those with
                  $\beta = 5.7$, were generated by the Berlin-Moscow-Adelaide group and the results published in
                  \cite{Bogolubsky:2009dc}. Note that in their paper, the lattice spacing was taken from $r_0$. The  Berlin-Moscow-Adelaide data was rescaled appropriately to follow our conventions.}
    \label{LatticeSetup}
   \end{center}
\end{table}

Our simulations are done at different lattice spacings. Therefore, in order to compare the propagators
computed at different $\beta$ values, the data has to be renormalized. In practice, we have renormalized the gluon propagator,
after performing a conic cut, by fitting the data to
\begin{eqnarray}
  D_{Lat} (p^2) ~ = ~ Z \, \frac{\left[ \ln \left(\frac{p^2}{\Lambda^2}\right) \right]^{-\gamma}}{p^2} \,,
  \label{uvfit}
\end{eqnarray}
Using the lowest order $\beta$-function and the coefficient of the lowest order anomalous gluon dimension for pure SU(3) Yang-Mills theory, extracted from e.g.~\cite{Gracey:2002yt,Chetyrkin:2004mf}, gives $\gamma = 13/22$. The fits to equation (\ref{uvfit}) were performed for a wide interval of momenta $\left[ p_{min} , p_{max} \right]$. For each lattice, the fitting range
was chosen so that $\chi^2/d.o.f. \sim 1$, while keeping the largest possible fitting interval. For the various lattices, the fitting range and the
quality of the fit can be found in TABLE \ref{UVFits}.\\

\begin{table}[t]
   \begin{center}
   \begin{tabular}{l@{\hspace{0.5cm}}c@{\hspace{0.3cm}}c@{\hspace{0.3cm}}c@{\hspace{0.3cm}}c@{\hspace{0.3cm}}c}
      \toprule
       \multicolumn{6}{c}{$\beta = 5.7$ \hspace{2cm} $a = 0.1838$ fm} \\
       \hline
      $L$                                   &    64            &  72           &  80              &  88              &  96 \\
      $p_{min}$  (GeV)       &  2.512        &   ---           & 2.471        & 2.486          & 2.498 \\
      $p_{max}$  (GeV)       &  4.418       &   ---           & 4.148         & 4.148         & 4.148 \\
      $\chi^2/d.o.f.$            &  1.65           &   ---           & 1.08           & 1.65            &  0.94  \\
      $Z_R$                          &  0.617(25) & 0.63(13)  & 0.621(29)  & 0.622(18)  & 0.64(31) \\
      \toprule
       \multicolumn{6}{c}{$\beta = 6.0$ \hspace{2cm} $a = 0.1016$ fm} \\
       \hline
      $L$                                    &    32             &  48             &  64                   &  80                 &   \\
      $p_{min}$  (GeV)       &  2.812         & 2.494         & 1.514              & 1.516             &  \\
      $p_{max}$  (GeV)      &  5.078         & 5.021         &  5.141             & 5.048             &  \\
      $\chi^2/d.o.f.$             & 0.91            & 0.97            &  0.89               & 1.20               &  \\
      $Z_R$                           & 0.149(21)  & 0.150(18)  &  0.1477(38)   &  0.1478(54)  &  \\
      \toprule
       \multicolumn{6}{c}{$\beta = 6.2$ \hspace{2cm} $a = 0.07261$ fm} \\
       \hline
      $L$                                      &   48               &  64                 &   &      &   \\
      $p_{min}$  (GeV)         & 2.121            & 1.591            &  &      &   \\
      $p_{max}$   (GeV)       &  5.286           & 5.110            &  &      &   \\
      $\chi^2/d.o.f.$               & 0.95              & 1.02              &  &      &   \\
      $Z_R$                             & 0.0743(72) &  0.0740(27)  &  &      &   \\
      \toprule
   \end{tabular}
   \caption{Ultraviolet fits to equation (\ref{uvfit}). Note that for $\beta = 5.7$ and the lattice $72^4$, the
                   $\chi^2/d.o.f.$ was never below 3. The renormalization constants $Z_R$ were computed
                   as described in the text and using $\mu = 3$ GeV as renormalization scale. The errors on $Z_R$ were
                   computed assuming Gaussian error propagation.}
    \label{UVFits}
   \end{center}
\end{table}

\begin{figure}[htbp]
   \centering
   \includegraphics[scale=0.4]{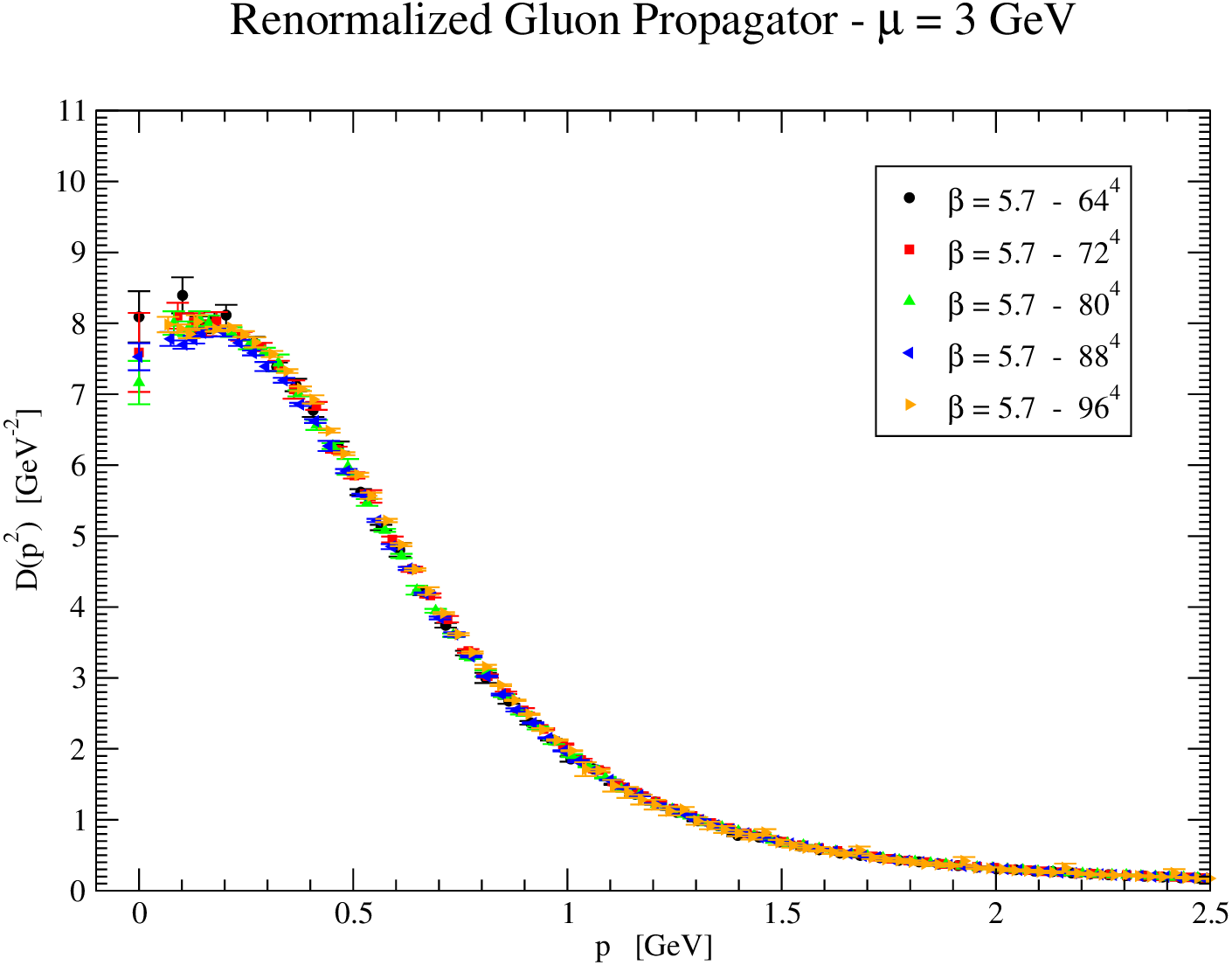}
   \caption{Renormalized gluon propagator for $\beta = 5.7$ simulations.}
   \label{FigPropBerlin}
\end{figure}

\begin{figure}[htbp]
   \centering
   \includegraphics[scale=0.4]{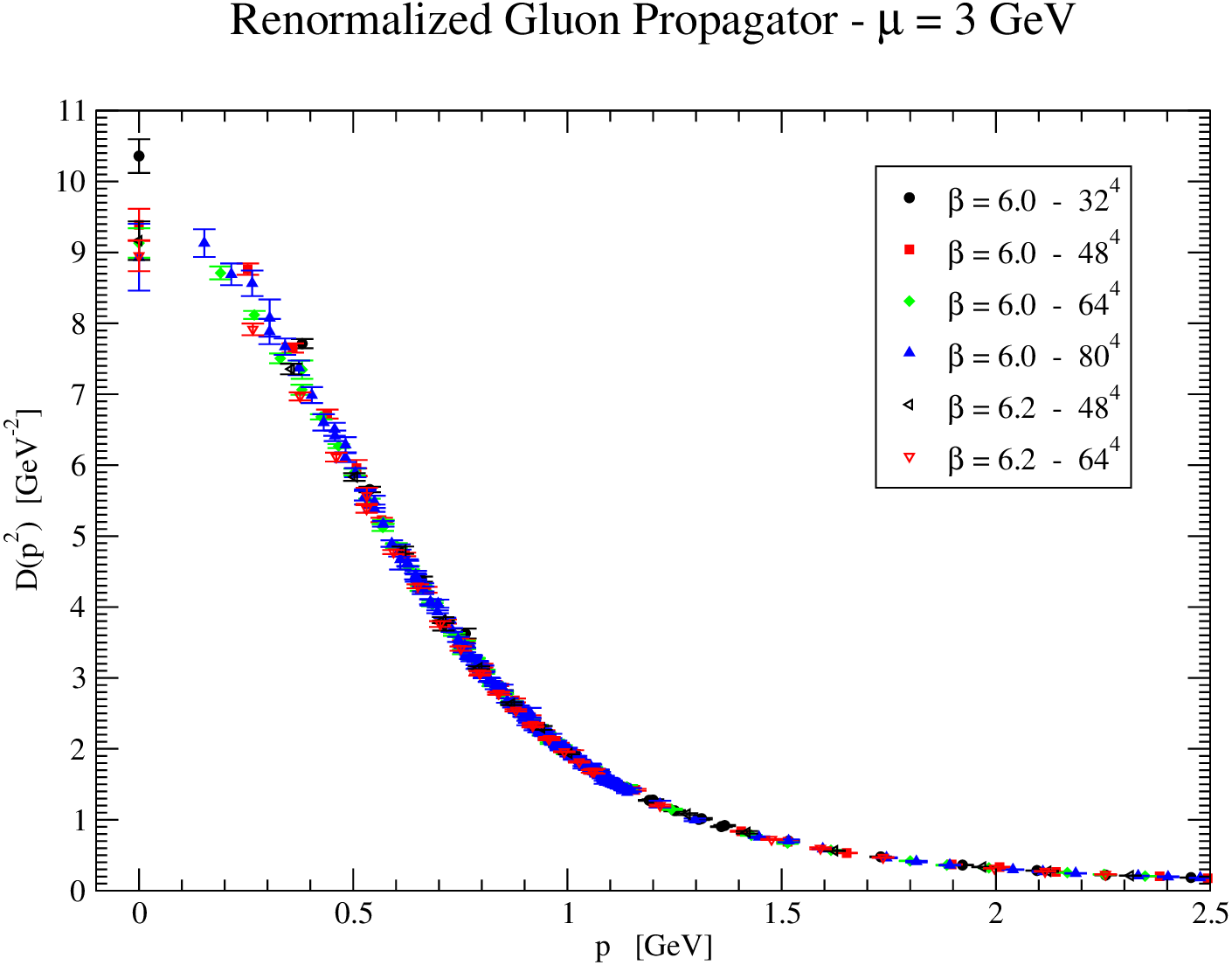}
   \caption{Renormalized gluon propagator for $\beta = 6.0$ and $\beta = 6.2$ simulations.}
   \label{FigPropCoimbra}
\end{figure}

The renormalized  gluon propagator,
\begin{equation}
  D(p^2) = Z_R \, D_{Lat} (p^2)\,,
\end{equation}
is related to the bare lattice propagator  $D_{Lat} (p^2)$ by requiring that
\begin{equation}
   \left.  D(p^2) \right|_{p^2 = \mu^2} ~ = ~ \frac{1}{\mu^2} \,,
   \label{renor}
\end{equation}
which defines (part of) a particular momentum subtraction (MOM) scheme. This condition defines the renormalization constant $Z_R$. Here we chose
$\mu = 3$ GeV. The values of $Z_R$ are reported in TABLE \ref{UVFits}. The renormalized propagator can be seen in FIG.~\ref{FigPropBerlin} for $\beta = 5.7$ and in FIG.~\ref{FigPropCoimbra} for the other $\beta$ values.\\

The $\beta = 5.7$ data seems to define a unique curve. In this sense, one can claim that
finite volume effects are under control. On the other hand, the propagators computed with larger  $\beta$ values show a small
dependence on the volume, especially in the infrared region. Note that, for the
largest two lattices, despite the
larger statistics, the $\beta = 5.7$ data displays a kind of ``wiggling'' structure. It is unclear whether this structure is of any relevance, as
the $\beta = 6.0$ and $\beta = 6.2$ data shows no such fluctuations in the infrared. Of course,
fluctuations in $D(p^2)$, even if they are small, can compromise the quality of the fittings.
We also observe that the $\beta = 5.7$ data is below the $\beta = 6.0$
and $\beta = 6.2$ data for momenta smaller than $\sim 400$ MeV, as it is clear
from FIG.~\ref{FigPropBerlin} and FIG.~\ref{FigPropCoimbra}.

\section{The lattice gluon propagator and the Refined Gribov-Zwanziger approach}
\subsection{Preliminaries}
In \cite{Dudal:2007cw,Dudal:2008sp} the gluon
propagator, among other things, was investigated by exploiting the Refined Gribov-Zwanziger action, and the tree level result (\ref{RZGtreelevel}) was derived. This propagator counts three
mass scales: (1) $M^2$, related to the condensation of the new fields $\left\{\varphi_{\mu}^{ab}, \overline\varphi_{\mu}^{ab}, \omega_{\mu}^{ab}, \overline\omega_{\mu}^{ab}\right\}$ introduced by Zwanziger
\cite{Zwanziger:1989mf,Zwanziger:1992qr} to localize the Gribov-Zwanziger action which was nonlocal after the first step of the construction, (2) $m^2$ related to the $\braket{A^2}$
condensate, and (3) $\gamma^4$ multiplying the horizon function, which is introduced to suppress the Gribov copies
in the functional integration. The differences between the (Refined) Gribov-Zwanziger action and the usual
Faddeev-Popov action can only show up in the infrared, as the difference between both is of a soft nature, i.e.~proportional to $\gamma^2$, see the action \eqref{GZ}. Indeed, if we formally set $\gamma^2=0$, the GZ action reduces to the usual Faddeev-Popov action as the $\left\{\varphi_{\mu}^{ab}, \overline\varphi_{\mu}^{ab}, \omega_{\mu}^{ab}, \overline\omega_{\mu}^{ab}\right\}$ fields can be integrated out to form a unity.\\

In principle, one can expect that the gluon propagator (\ref{RZGtreelevel}) should be able to reproduce the lattice data in a certain momentum
region.  Note that being a tree level result, it does
not include the logarithmic correction which has been observed, for example, when fitting the ultraviolet
region. Remember that the logarithmic dependence was explored to renormalize the lattice gluon propagator.
Therefore, assuming that (\ref{RZGtreelevel}) describes the lattice data, one can expect that it will not reproduce
the ultraviolet data, the difference being caused by the lack of the perturbative logarithmic correction. Anyway, one can
explore the lattice results to check if (\ref{RZGtreelevel}) can reproduce the propagators reported in
FIGS.~\ref{FigPropBerlin} and ~\ref{FigPropCoimbra} up to a certain maximum momentum, as in the infrared, we expect that the logarithm will ``freeze'' due the presence of infrared mass scales. Furthermore, given
the relation between the different mass scales and the condensates, by setting either $M^2$ or $m^2$ to zero
one can check for the corresponding contribution to nonperturbative physics.

\subsection{Gluon propagator and evidence for the $d=2$ gluon condensate $\braket{A^2}$}
Let us first consider the condensation of the extra fields $\left\{\varphi_{\mu}^{ab}, \overline\varphi_{\mu}^{ab}, \omega_{\mu}^{ab}, \overline\omega_{\mu}^{ab}\right\}$. Given that the lattice gluon propagator does
not vanish at zero momentum, one must have $M^2 \neq 0$. Indeed, the motivation to introduce the
condensate associated with the new ghost-type fields was precisely to be able to have a $D(0) \ne 0$
\cite{Dudal:2007cw,Dudal:2008sp}.
We shall need the following correspondence between the tree level gluon mass $m^2$ and the condensate $\braket{A^2}$ \cite{Verschelde:2001ia,Dudal:2005na}
\begin{equation}\label{corr}
    \braket{g^2A^2}= -\zeta_0 m^2\,,\qquad \zeta_0=\frac{9}{13}\frac{N^2-1}{N}\,,
\end{equation}
which follows from the construction of \cite{Verschelde:2001ia}. From this relation, it is clear that the presence of the condensate requires a nonvanishing $m^2$. This can be tested setting $m^2 = 0$ in
the tree level expression (\ref{RZGtreelevel}) and trying to fit the lattice data to
\begin{equation}
   D(p^2) = \frac{p^2 + M^2}{p^4 + M^2 p^2 + 2 g^2 N \gamma^4 } \,.
   \label{NoA2}
\end{equation}
Despite the similar structure of (\ref{RZGtreelevel}) and (\ref{NoA2}), the lattice data distinguishes quite clearly
the two functional forms. Indeed, while (\ref{RZGtreelevel}) is able to reproduce the lattice propagator on a
wide range of momentum starting at 0 GeV and going up to $1 - 1.5$ GeV, in the sense that the
corresponding fit have $\chi^2 /d.o.f. < 2$,  the fits corresponding to (\ref{NoA2}) always have a $\chi^2 /d.o.f.$ larger than three, and can as such be rejected.\\

\begin{figure}[htbp]
   \centering
   \includegraphics[scale=0.4]{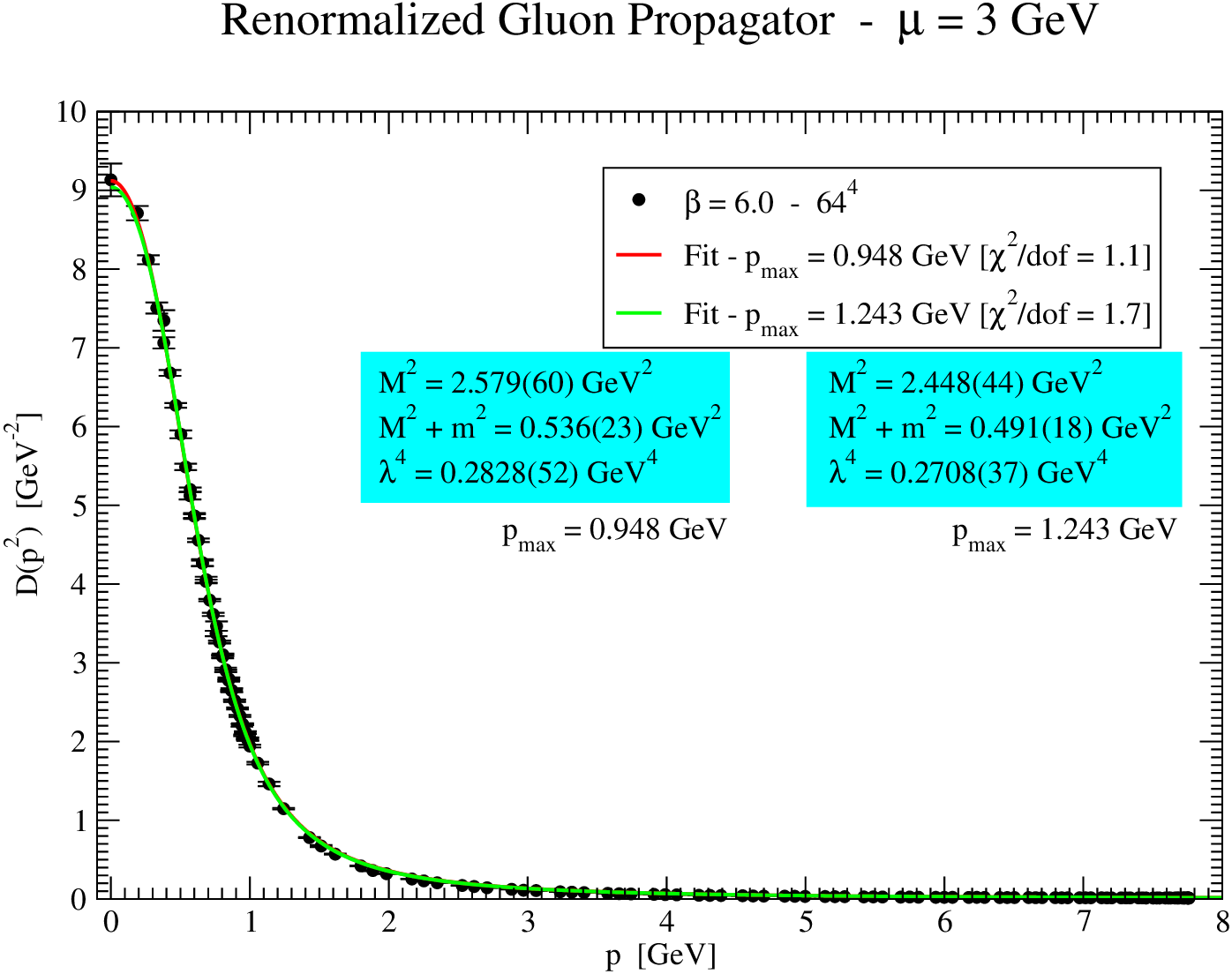}
   \caption{Gluon propagator and fit to (\ref{RZGtreelevel}) using the momentum range $[0 , p_{max} ]$.
                   $p_{max} = 1.243$ GeV is the largest fitting range which has a $\chi^2/d.o.f. < 2$. The figure
                   includes the outcome of the fits for the two fitting ranges considered.}
   \label{FullPropFit}
\end{figure}

\begin{figure}[htbp]
   \centering
   \includegraphics[scale=0.4]{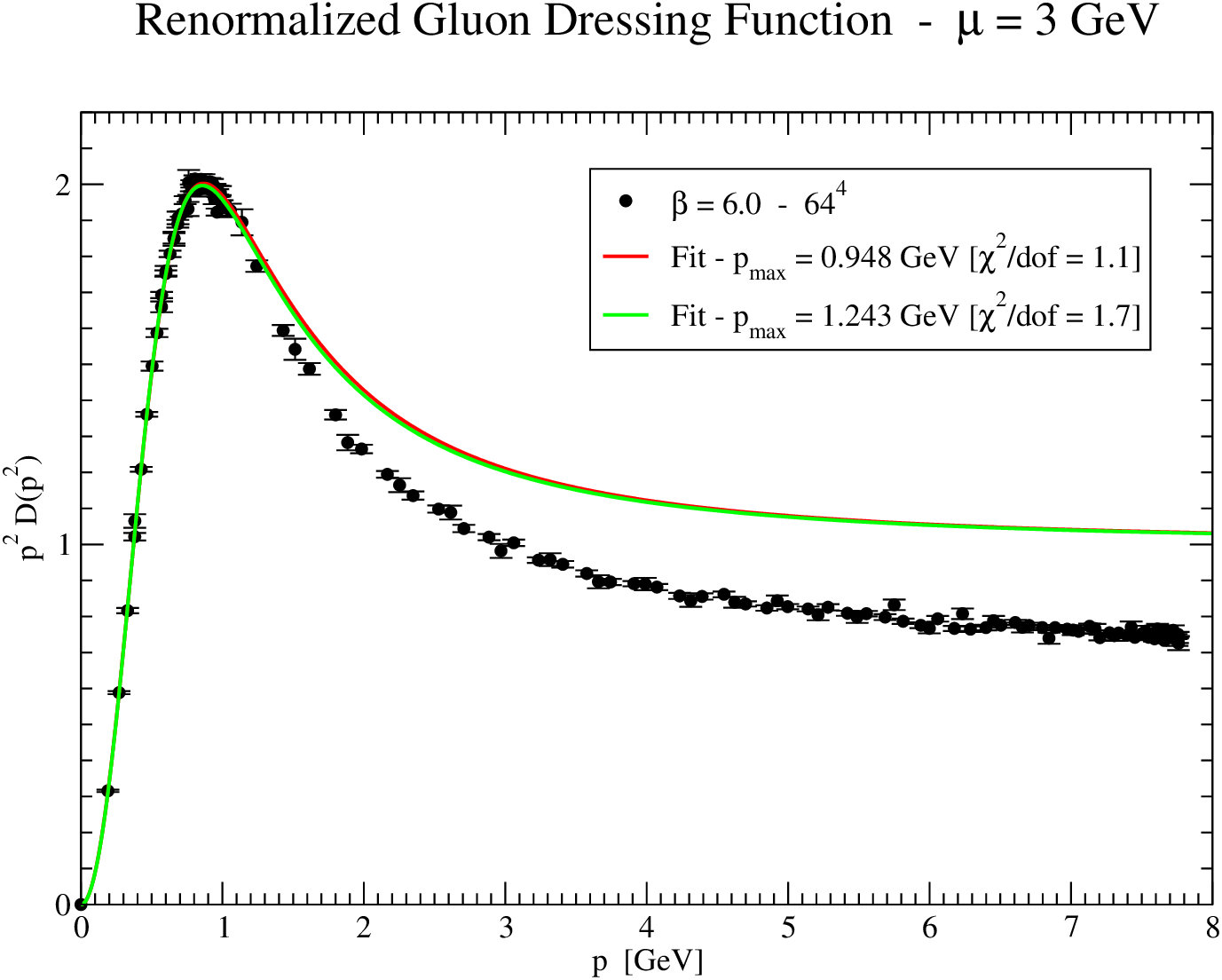}
      \caption{The same as in FIG.~\ref{FullPropFit} but for the gluon dressing function $p^2 D(p^2)$. The dressing function
                    provides a clear picture of the differences between (\ref{RZGtreelevel}) and the lattice data in the
                    ultraviolet region.}
   \label{FullDressFit}
\end{figure}

As an example of a fit with $m^2\neq0$, in FIG.~\ref{FullPropFit} we show the renormalized gluon propagator computed using the
$\beta = 6.0$ and $64^4$ lattice and the fits corresponding  to (\ref{RZGtreelevel}).
Although the fits use only the momentum in $[0 , p_{max} ]$, in FIG.~\ref{FullPropFit} we show the propagator if one uses
(\ref{RZGtreelevel}) over the entire momentum region. There is a small difference between the lattice data and
the prediction of (\ref{RZGtreelevel}) in the ultraviolet region which is clearly seen in the gluon
dressing function - see FIG.~\ref{FullDressFit}. As discussed previously, the small observed
differences\footnote{For the highest lattice momenta $p = 7.76$ GeV, the measured propagator is
0.01205(32) GeV$^{-2}$, while (\ref{RZGtreelevel}) predicts 0.0172 GeV$^{-2}$.}
are expected as (\ref{RZGtreelevel}) does not take into account the perturbative logarithmic corrections.

Our interpretation of the fits to (\ref{RZGtreelevel}) and (\ref{NoA2}) is that the lattice data points towards
a nonvanishing gluon condensate $\braket{A^2}$. In Section III.D, we shall discuss this in more detail and extract an estimate for $\braket{A^2}$.

\subsection{Measuring the scales in the Refined Gribov-Zwanziger gluon propagator using the lattice data}
\begin{figure}[t]
   \centering
   \includegraphics[scale=0.4]{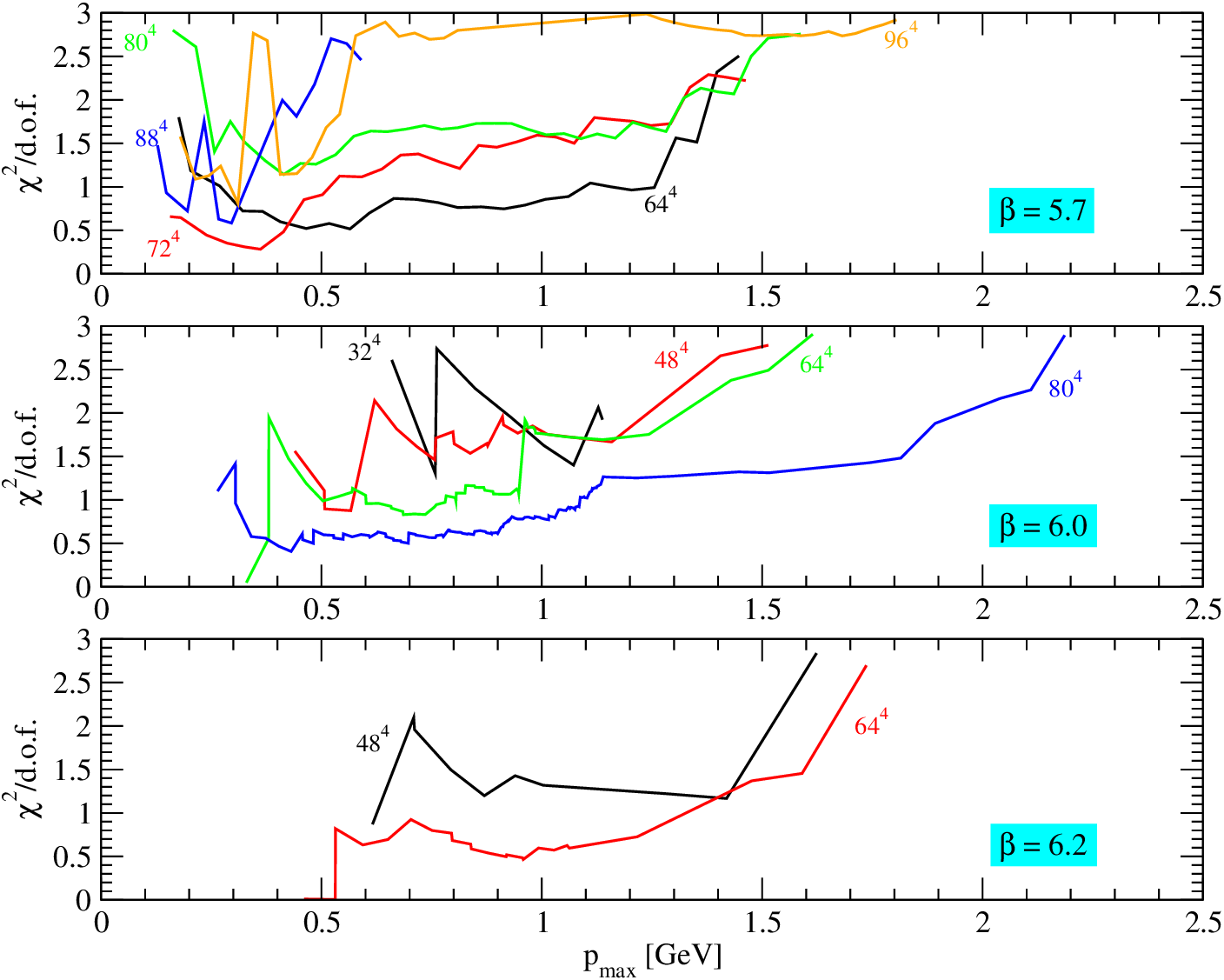}
   \caption{Fitting the propagator to (\ref{RZGtreelevel}): $\chi^2/d.o.f.$ as a function of the maximum fitting
                  momenta $p_{max}$ for each lattice.}
   \label{Figchi2Pmax}
\end{figure}

In this section we aim to investigate the compatibility of the tree level gluon propagator computed using the Refined Gribov-Zwanziger action and the lattice data. In particular, we would like to measure the different
mass parameters in (\ref{RZGtreelevel}). As discussed previously, it is not expected that (\ref{RZGtreelevel})
is able to describe the lattice propagator for the full range of momenta.  Therefore, we perform a sliding
window analysis, i.e.~we shall fit the propagator using momenta in $[0 , p]$, with increasing values for $p$. Then,
the $\chi^2/d.o.f.$ can be used to establish a maximum range of momenta described by
(\ref{RZGtreelevel}) - see FIG.~\ref{Figchi2Pmax}. For the largest two $\beta$ values and for the
largest lattices, the Refined Gribov-Zwanziger tree level propagator is able to describe the lattice data well
above 1 GeV. In particular, for the largest volume, being the $\beta = 6.0$ and $80^4$ case, the lattice gluon
propagator can be fitted by (\ref{RZGtreelevel}) beyond 1.5 GeV. We draw the reader's attention by noticing that
for this particular set of data the ``perturbatively'' inspired expression (\ref{uvfit}) describes the lattice data
starting from 1.5 GeV (see TABLE \ref{UVFits}). For the smaller $\beta = 5.7$ simulations,
the situation is similar, with the exception of the largest two lattices ($88^4$ and $96^4$). In the Appendix, we have spent a few words about these latter two lattices, and we motivate why we shall keep them out of our analysis. We shall however use the other $\beta = 5.7$ data to check our results later on.\\

\begin{figure}[t]
   \centering
   \includegraphics[scale=0.4]{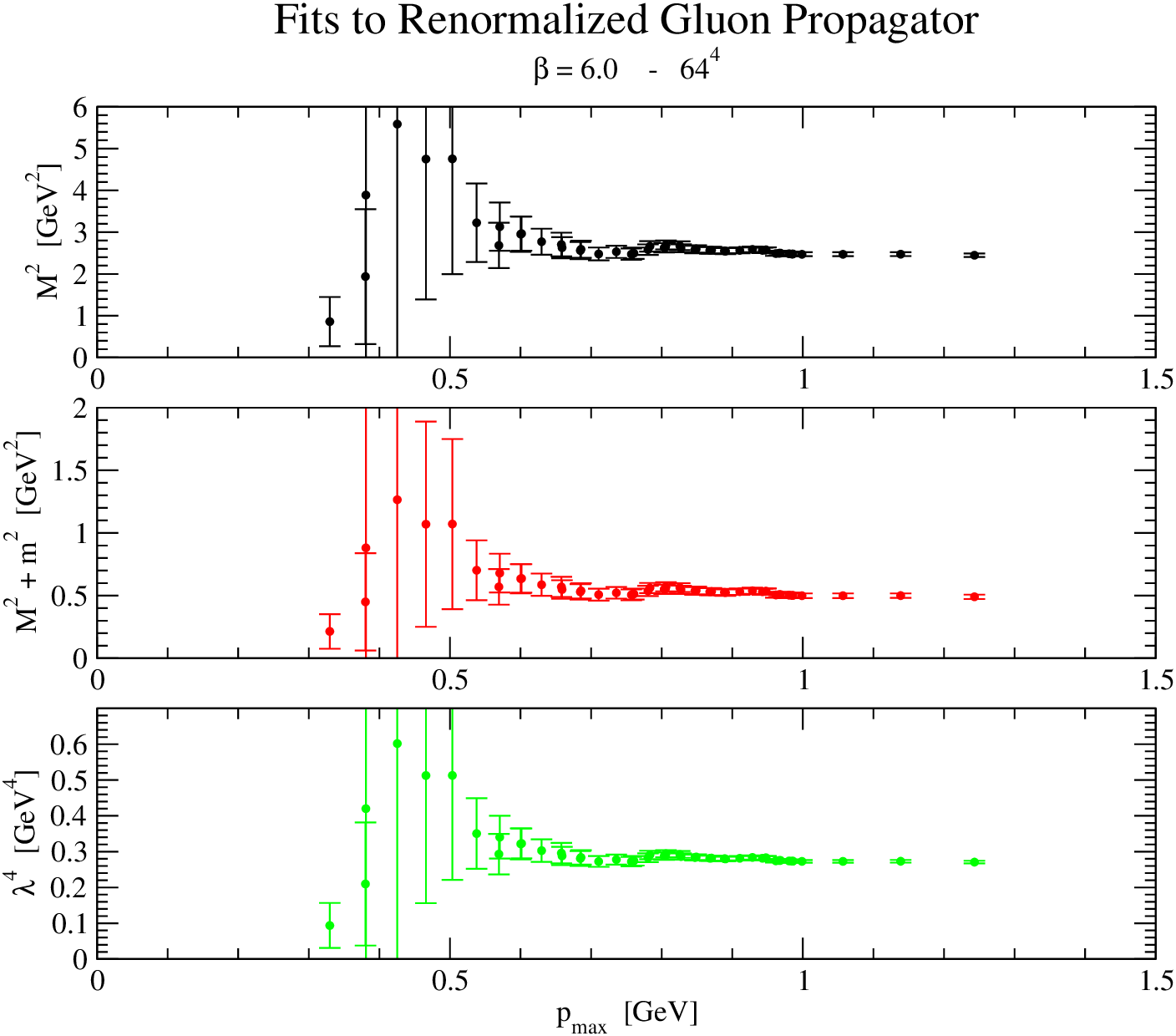}
   \caption{Evolution of the fitting parameters with $p_{max}$ for $\beta = 6.0$ and $64^4$ data.}
   \label{Fitspmax64}
\end{figure}

In FIG.\ref{Fitspmax64} we report the result of fitting (\ref{RZGtreelevel}) to the renormalized gluon propagator
computed from the $\beta = 6.0$ and $64^4$ lattice data as a function of the
fitting range $[ 0 , p_{max}]$. Similar plots can be shown for the remaining fits.
As FIG.\ref{Fitspmax64} shows, the estimated values for $M^2$, $M^2 + m^2$ and
$\lambda^4 = 2 g^2 N \gamma^4 + M^2m^2$ are stable against a change on $p_{max}$. For each simulation,
as a set of values, we choose those which correspond to the largest fitting range with a $\chi^2/d.o.f. \sim 1$.
For example, for the $\beta = 6.0$ and $64^4$ data, we take $p_{max} = 0.929$ GeV and
$M^2 =  2.589 \pm 0.068$ GeV$^2$, $M^2 + m^2 =  0.539 \pm 0.025$ GeV$^2$,
$\lambda^4 =   0.2837   \pm 0.0059$ for a $\chi^2 /d.o.f. =  1.07$.
When the $\chi^2/d.o.f.$ never crosses or become to close to 1, such as happens in the smallest
fitting lattice volume, we choose the set of values which minimizes $\chi^2/d.o.f.$ for the largest possible
fitting range.\\

\begin{figure}[htbp]
\vspace*{0.5cm}
   \centering
   \includegraphics[scale=0.4]{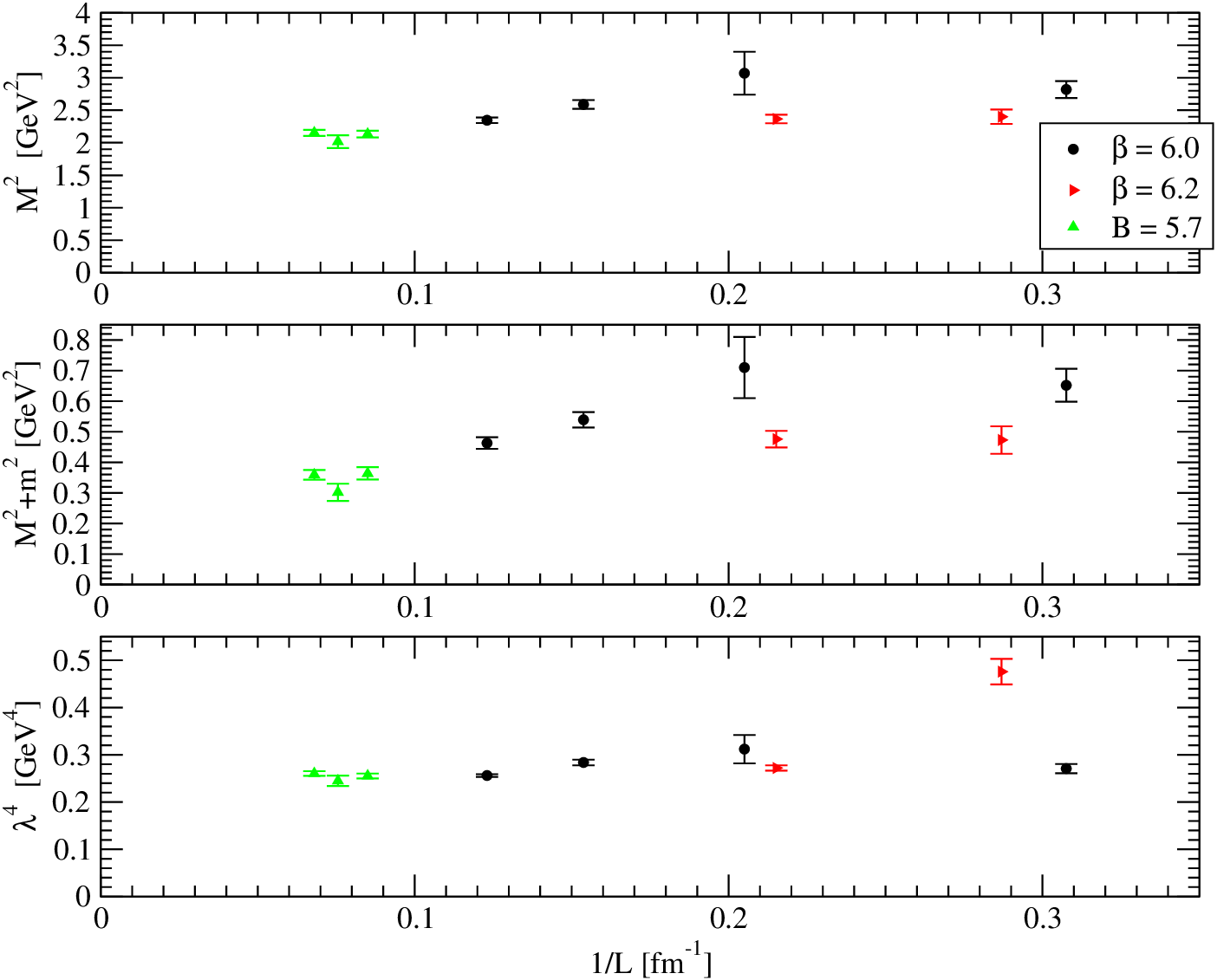}
   \caption{Parameters for the tree level gluon propagator of the Refined Gribov-Zwanziger action, computed fitting
                  the renormalized gluon propagator, as a function of the inverse of the lattice length $L$. The reader
                  should remember that, for $\beta = 5.7$ and $72^4$, the lattice data was not well described by the UV fit (\ref{uvfit}) used to define
                  the renormalization constant $Z_R$, see the discussion on the renormalization procedure. This can explain the observed fluctuations
                  in the $\beta = 5.7$ results.}
   \label{FigFitsfinais}
\end{figure}

In TABLE \ref{Fitsfinais} we report the estimates of the different parameters defining the Refined Gribov-Zwanziger
tree level gluon propagator for each lattice simulation. The values are plotted in FIG.~\ref{FigFitsfinais} as a function of the inverse of the lattice length $L$. The data shows a small dependence on $1/L$, especially for $M^2+m^2$, and on the lattice
spacing, i.e.~on $\beta$. Nevertheless, for $\beta = 6.0$, the four volumes can be combined to perform a linear
extrapolation to the infinite volume limit.\\

\begin{table}[htbp]
   \centering
   \begin{tabular}{l@{\hspace{0.7cm}} l@{\hspace{0.5cm}}l@{\hspace{0.5cm}}l@{\hspace{0.5cm}}l@{\hspace{0.5cm}}l@{\hspace{0.5cm}}l}
      \toprule
      $L$    &  $p_{max}$  &  $M^2$                        & $M^2+m^2$                 & $\lambda^4$               &  $\chi^2 / d .o.f.$ \\
      \toprule
      \multicolumn{6}{c}{$\beta = 5.7$} \\
     64   &  1.255          &  $2.132  \pm  0.052$  &   $0.364 \pm  0.020$  &  $0.2553 \pm 0.0051$  &  0.99  \\
     72   &   0.814          &  $2.017 \pm  0.097$  &   $0.302 \pm  0.028$  &  $0.245    \pm  0.011$   & 1.21  \\
     80   &   1.089          &  $2.151  \pm  0.047$ &   $0.359 \pm  0.016$  &  $0.2604 \pm 0.0049$  & 1.55  \\
      \toprule
      \multicolumn{6}{c}{$\beta = 6.0$} \\
      32  &    1.072         &  $2.82 \pm 0.13$       &    $0.652 \pm 0.054$  &  $0.2708  \pm 0.0096$ &   1.40  \\
      48  &    0.757         &   $3.07  \pm 0.33$     &    $0.71   \pm 0.10$     &  $0.312 \pm 0.030$        &  1.46  \\
      64  &    0.929         &   $2.589 \pm 0.068$  &   $0.539 \pm 0.025$   &  $0.2837  \pm 0.0059$  &  1.07  \\
      80  &    1.103         &   $2.346 \pm 0.043$  &   $0.463 \pm 0.019$   &  $0.2561  \pm 0.0030$  &  1.03  \\
      \toprule
      \multicolumn{6}{c}{$\beta = 6.2$} \\
     48   &    1.419         &   $2.40  \pm 0.11$      & $0.473 \pm 0.045$     &  $0.2677 \pm 0.0095 $  &  1.17 \\
     64   &    1.476         &   $2.366 \pm 0.066$   & $0.476 \pm 0.027$    &   $0.2721 \pm- 0.0057$  &  1.37 \\
         \toprule
   \end{tabular}
   \caption{Tree level gluon propagator parameters from fitting the Refined Gribov-Zwanziger propagator \eqref{RZGtreelevel} to the renormalized lattice gluon propagator.
                 The errors reported are statistical and computed assuming Gaussian error propagation. }
   \label{Fitsfinais}
\end{table}

As a function of $1/L$, $M^2$ is reasonably  well described by a linear function. Indeed, the $\chi^2/d.o.f.$
of the fit is 2.13, giving
\begin{equation}\label{scale1}
M^2 = 2.15 \pm 0.13~\text{GeV}^2\,,
\end{equation}
in good agreement with the value computed from
the largest $\beta = 5.7$ volume. \\

For $M^2 + m^2$, the linear fit gives an infinite volume value of
\begin{equation}\label{scale2}
M^2+m^2 = 0.337 \pm   0.047~\text{GeV}^2\,,
\end{equation}
for a $\chi^2/d.o.f. = 2.04$.\\

\begin{figure}[t]
   \centering
   \includegraphics[scale=0.4]{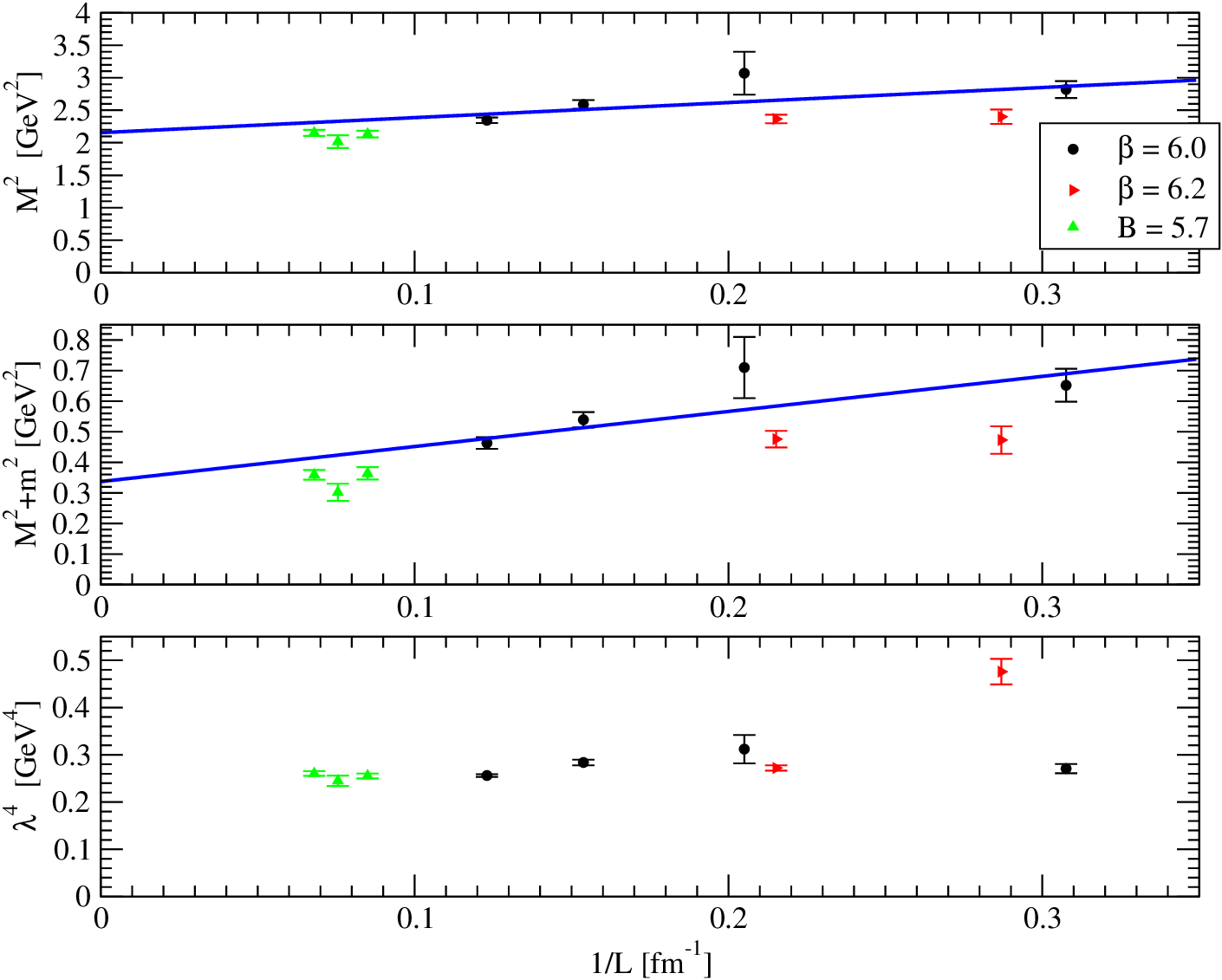}
   \caption{The same as FIG.~\ref{FigFitsfinais} but including the linear extrapolations for $M^2$ and
                   $M^2 + m^2$, which are obtained using the $\beta=6.0$ data. The large volume $\beta=5.7$ data serves as a consistency check, as explained before in the text.}
   \label{FigFitsfinaisFits}
\end{figure}

For $\lambda^4$, the linear extrapolation
has a $\chi^2/d.o.f.$ larger than 3. Fortunately, it seems that $\lambda^4$ shows the smallest dependence on
$1/L$ and the lattice spacing, with the largest volumes providing numbers which are compatible, within one
standard deviation. Therefore, given the results reported in TABLE \ref{Fitsfinais}
for the largest volumes, one can claim that
\begin{equation}\label{scale3}
\lambda^4 = 0.26~\text{GeV}^4\,,
\end{equation}
which are the reliable digits from the largest two lattices -- see the TABLE. The linear extrapolations can
be see in FIG.~\ref{FigFitsfinaisFits}. We observe that the figures for the $\beta = 5.7$ data and the linearly extrapolated results are pretty close, giving us further confidence in the extrapolation.\\

If one uses the extrapolated values, one can write down that
\begin{equation}\label{scale4}
m^2 = - 1.81 \pm 0.14~\text{GeV}^2\,.
\end{equation}
Simultaneously we find
\begin{equation}
2 g^2 N \gamma^4 = 4.16 \pm 0.38~\text{GeV}^4\,.
\end{equation}
Furthermore, assuming that (\ref{RZGtreelevel})
describes the infrared gluon propagator, then
\begin{equation}
   D(0) = \frac{M^2}{\lambda^4} = 8.3 \pm 0.5 ~\text{GeV}^{-2}\,.
\end{equation}
The zero momentum gluon propagator computed using the extrapolated values for $M^2$ and $\lambda^4$
is in excellent agreement, within one standard deviation,
with the lattice $D(0)$ computed from lattice QCD for $\beta = 5.7$ where $D(0) \sim 7 - 8.5$ GeV$^{-2}$,
$\beta = 6.0$ and $80^4$ data where $D(0) = 8.93 \pm 0.47$ GeV$^{-2}$
and for $\beta = 6.2$ and $64^4$ data which has a $D(0) = 8.95 \pm 0.22$ GeV$^{-2}$.

\subsection{Extracting a value for the dimension two gluon condensate $\braket{g^2A^2}$}
In order to obtain an estimate that can be compared with other values available on the market, we shall rely on the renormalization group. In particular, we wish to compare with the values \eqref{A3} and \eqref{A3bis}, being
\begin{equation}\label{A3nog}
\braket{g^2A^2} =5.1^{+0.7}_{-1.1}~\text{GeV}^2\,,
\end{equation}
and
\begin{equation}\label{A3bisnog}
\braket{g^2A^2} =4.4\pm0.4~\text{GeV}^2\,.
\end{equation}
For the relevant one loop renormalization group equations, we have, in any (massless) renormalization scheme\footnote{We recall that the lowest order anomalous dimensions are universal quantities.} \cite{Verschelde:2001ia,Dudal:2005na}
\begin{eqnarray}\label{corr2}
    \mu\frac{\p}{\p \mu} g^2&=& -2\beta_0g^4\,,\qquad \beta_0=\frac{11}{3}\frac{N}{16\pi^2}\,,\nonumber\\
        \mu\frac{\p}{\p \mu} m^2&=& \gamma_0 g^2m^2\,,\qquad \gamma_0=-\frac{3}{2}\frac{N}{16\pi^2}\,.
\end{eqnarray}
Hence, our estimate \eqref{scale4} corresponds to a positive gluon condensate, as using \eqref{corr} yields for $N=3$
\begin{equation}\label{A2}
\braket{g^2A^2} = 3.35 \pm 0.26~\text{GeV}^2\,.
\end{equation}
i.e.~a positive gluon condensate. We recall that in this work we have renormalized at a scale $\mu=3$ GeV.\\

The value \eqref{A3nog} was obtained in the so-called $T$-scheme, which is kind of MOM scheme compatible with the renormalization prescription \eqref{renor}, at a renormalization scale $\mu=10$ GeV. The fundamental scale $\Lambda_T$ of this $T$-scheme is related to the  conventional $\MSbar$ one through the conversion formula \cite{Boucaud:2008gn}
\begin{equation}\label{A4}
\Lambda_T=\lms e^{507/792}\,.
\end{equation}
Using \eqref{corr2}, we have at one loop
\begin{equation}\label{A5}
    \mu\frac{\p}{\p \mu} m^2 = \frac{\gamma_0}{2\beta_0} \frac{1}{\ln\frac{\mu}{\Lambda_T}}m^2\,.
\end{equation}
Introducing the auxiliary variable $\xi=\ln\frac{\mu}{\Lambda_T}$, this can be easily integrated to
\begin{equation}\label{A6}
    m^2=m_0^2 \left(\frac{\xi}{\xi_0}\right)^{\frac{\gamma_0}{2\beta_0}}=m_0^2\left(\frac{\ln\frac{\mu}{\Lambda_T}}{\ln\frac{\mu_0}{\Lambda_T}}\right)^{-9/44}\,,
\end{equation}
using the numbers given in \eqref{corr2}. The estimate $\lms=0.224$ GeV determined in \cite{Boucaud:2008gn} consequently leads to
\begin{equation}\label{A6}
    \braket{g^2A^2}^{\mu=10\,\mathrm{GeV}}= 3.03 \pm 0.24~\text{GeV}^2\,,
\end{equation}
employing \eqref{corr} and $\braket{g^2A^2}= 3.29$ GeV at $\mu_0=3$ GeV as input values. We notice that our estimate is at least in the same ballpark as the ones of \eqref{A3nog} and \eqref{A3bisnog}, which were obtained in a completely independent way. In these works, it was observed that even at relatively large
momenta $Q^2$, there was a discrepancy between the perturbatively expected results, and the lattice estimates for the gluon or ghost propagator and
strong running coupling constant. Usually, such discrepancies can be accommodated for by nonperturbative power corrections. It was discussed in \cite{Boucaud:2001st,Boucaud:2008gn} that a power correction proportional to $\braket{A^2}/Q^2$ was necessary to obtain a sensible estimate for
e.g. $\lms$. In the current work, we obtained a lattice estimate for the same condensate $\braket{g^2A^2}$ in a completely different way, hence it is
quite remarkable a compatible value is retrieved at the end of each analysis.\\

\subsection{Extracting an infrared mass scale from the gluon propagator}
As a final effort we would like to estimate an infrared mass scale by using the gluon propagator. A similar attempt was done in \cite{Bornyakov:2009ug}. The infrared lattice data is well described by equation
(\ref{RZGtreelevel}), which however depends on multiple mass scales. For small enough momenta, $p\lessapprox 0.2$ GeV, the propagator (\ref{RZGtreelevel}) is well approximated by the so-called pole (or Yukawa) propagator fit
\begin{equation}
   D(p^2)  \approx \frac{M^2}{ \left(M^2 + m^2\right) p^2 + \lambda^4  } = \frac{Z}{ p^2 +m^2_{IR}  } \,,
   \label{RZGtreelevelSmallq}
\end{equation}
where
\begin{equation}\label{infraredmass}
        m^2_{IR} = \frac{\lambda^4}{M^2 + m^2}
\end{equation}
is the infrared mass scale we can associate to infrared pure QCD. Using the infinite volume estimates for $\lambda^4$ and $M^2 + m^2$, it follows that
\begin{equation}
m_{IR} = 771(108)~\text{MeV}\,.
\end{equation}
This infrared mass scale is in excellent agreement with the infrared mass scale estimates from large volume SU(3) simulations
\cite{Oliveira:2009nn}, where a gluon mass in the range 600 - 800 MeV was claimed,
and in good agreement with the recent value obtained in \cite{Oliveira:2010xc}, where a $m_{IR} = 651(12)~\text{MeV}$ was measured. Furthermore, the value given in equation \eqref{infraredmass} agrees also well with the
SU(2) result found in \cite{Burgio:2009xp}, i.e. $m_{IR} = 856(8)$ MeV, and it is only slightly larger than the SU(2) infrared mass scale derived in
\cite{Bornyakov:2009ug}. In the latter work, it was however noticed that a pole fit like \eqref{RZGtreelevelSmallq} does not work out well. Indeed, we observe that in our case, the fitting range is only something like $p\in[0,0.2]$ working in GeV, while the corresponding mass is about $0.85$ GeV, so the name of a ``pole propagator fit'' is a bit misguided of course as at $p\approx m_{IR}$ the fit is already invalid. As an alternative, the authors of \cite{Bornyakov:2009ug} proposed a Gaussian fit in the continuum
\begin{equation}\label{altfit}
   D(p^2) = B e^{-(p-p_0)^2/m^2_{IR}}\,,
\end{equation}
identifying from this an infrared mass scale $m^2_{IR}$. As far as we know, there is no theoretical motivation behind this kind of propagator yet. Roughly said, one should identify a mechanism that can generate a momentum-exponential into the effective action in the $A-A$ sector. It should also be noticed that in order to write down the expression \eqref{altfit}, an external momentum vector $p_0$ must be introduced, thereby sacrificing Lorentz invariance\footnote{Or more precisely, rotational invariance as we are working in Euclidean space. The breaking is evident as the resulting propagator is no longer a function of the invariant~$p^2$.}.

\section{Conclusions}
In this paper, we have shown that
\begin{equation}\label{conc1}
   D(p^2) = \frac{p^2 + M^2}{p^4 + \left(M^2 + m^2\right) p^2 + 2 g^2 N \gamma^4 + M^2 m^2  }\,,
\end{equation}
which is the analytical tree level version of the gluon propagator found in the (Refined) Gribov-Zwanziger formalism \cite{Dudal:2008sp}, can describe very well the lattice data for the SU(3) Landau gauge gluon propagator in the infrared. More precisely, for momenta up to 1.5 GeV, a good fit was established, which was only possible with nonzero values for all mass parameters appearing in \eqref{conc1}. We discussed their continuum extrapolation, which yielded the following estimates:
\begin{equation}\label{conc2}
M^2~=~2.15\pm0.13~\text{GeV}^2\,,\qquad m^2~=~-1.81\pm0.14~\text{GeV}^2\,,\qquad 2g^2N\gamma^4~=~4.16\pm0.38~\text{GeV}^4\,,
\end{equation}
giving the following continuum value for $D(0)$,
\begin{equation}\label{conc3}
D(0)~=~8.3\pm0.5~\text{GeV}^{-2}\,,
\end{equation}
which is in good agreement with large volume lattice data.\\

Since $m^2$ is related to the $\braket{A^2}$ condensate, we were also able to present the value
\begin{equation}\label{conc4}
\braket{g^2A^2}^{\mu=10~\text{GeV}} ~=~ 3.03 \pm 0.24~\textrm{GeV}^2\,,
\end{equation}
which compares fairly with other estimates of this $d=2$ gluon condensate.\\

We conclude that the current work has collected evidence that the Refined Gribov-Zwanziger formalism is perfectly well capable of explaining the infrared behaviour of the (lattice) Landau gauge gluon propagator, with its nonvanishing zero momentum limit. This is good news, as fitting lattice data is one thing, but one should also be able to explain which effects are behind a particular fit. We notice that also certain Schwinger-Dyson results for the same propagator describe the lattice data well, see e.g.~\cite{Fischer:2008uz,Binosi:2009qm}, perhaps indicative of a close connection between these formalisms and the (R)GZ one, a fact already explored in the work \cite{Huber:2009tx}. At the same time, we have also provided further evidence that one cannot ignore the effects of the dimension two gluon condensate $\braket{A^2}$ in the Landau gauge.

\appendix
\section{A few words about the $\beta = 5.7$ data at volumes $88^4$ and $96^4$}
\begin{figure}[htbp]
   \centering
   \includegraphics[scale=0.4]{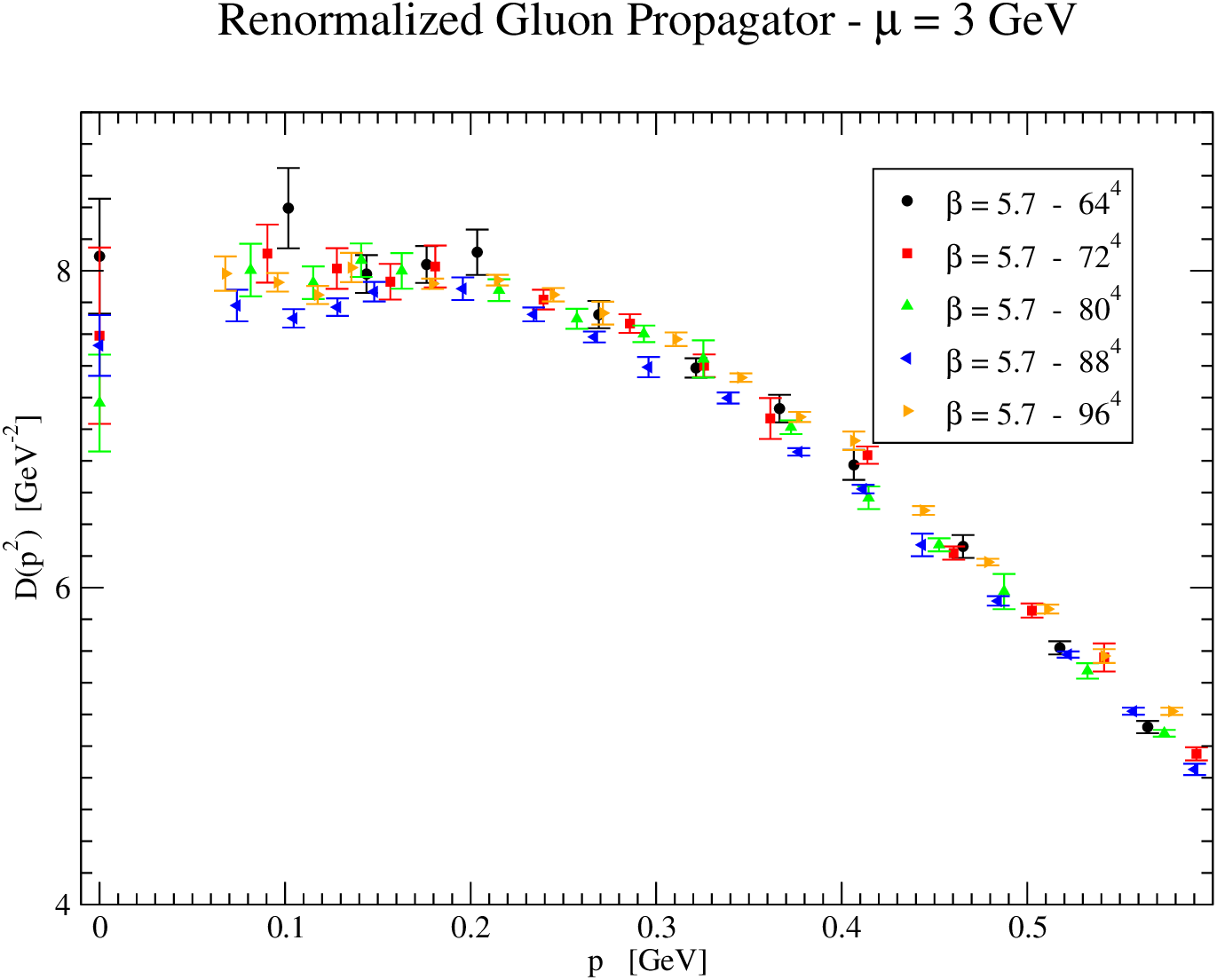}
      \caption{A zoom of the infrared gluon propagator computed at $\beta = 5.7$.}
   \label{BerlinGlueIR}
\end{figure}

The gluon propagator computed at $\beta = 5.7$ and volumes $88^4$ and $96^4$ turns out to be rather problematic to fit. A closer look shows that for these largest two
lattices the data fluctuates quite strong in the infrared, see FIG. \ref{BerlinGlueIR}. If one ignores these data points, the data does behave similarly as for the $\beta = 6.0$
and $\beta = 6.2$
simulations. For example, for the largest volume, removing the smallest five momenta,
i.e.~taking into account only $p \ge 214$ MeV, the largest fitting range with a $\chi^2/d.o.f < 1.6$ corresponds to a $p_{max} = 1.587$ GeV. \\

The observed infrared fluctuations can be understood from the combination of the (poor) statistics and the particular choice of $\beta$.\\

\begin{figure}[htbp]
   \centering
   \includegraphics[scale=0.3]{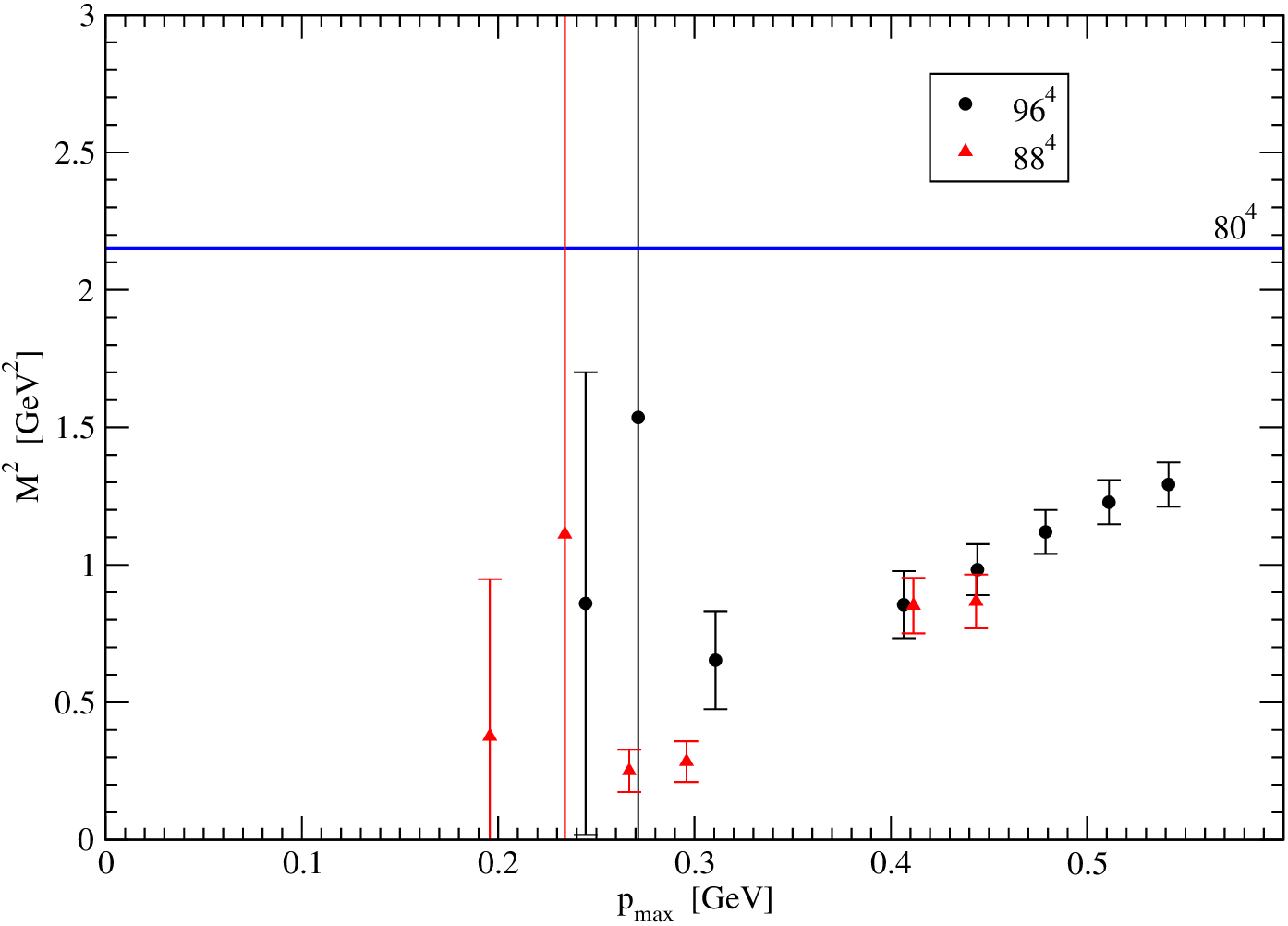} \qquad
   \includegraphics[scale=0.3]{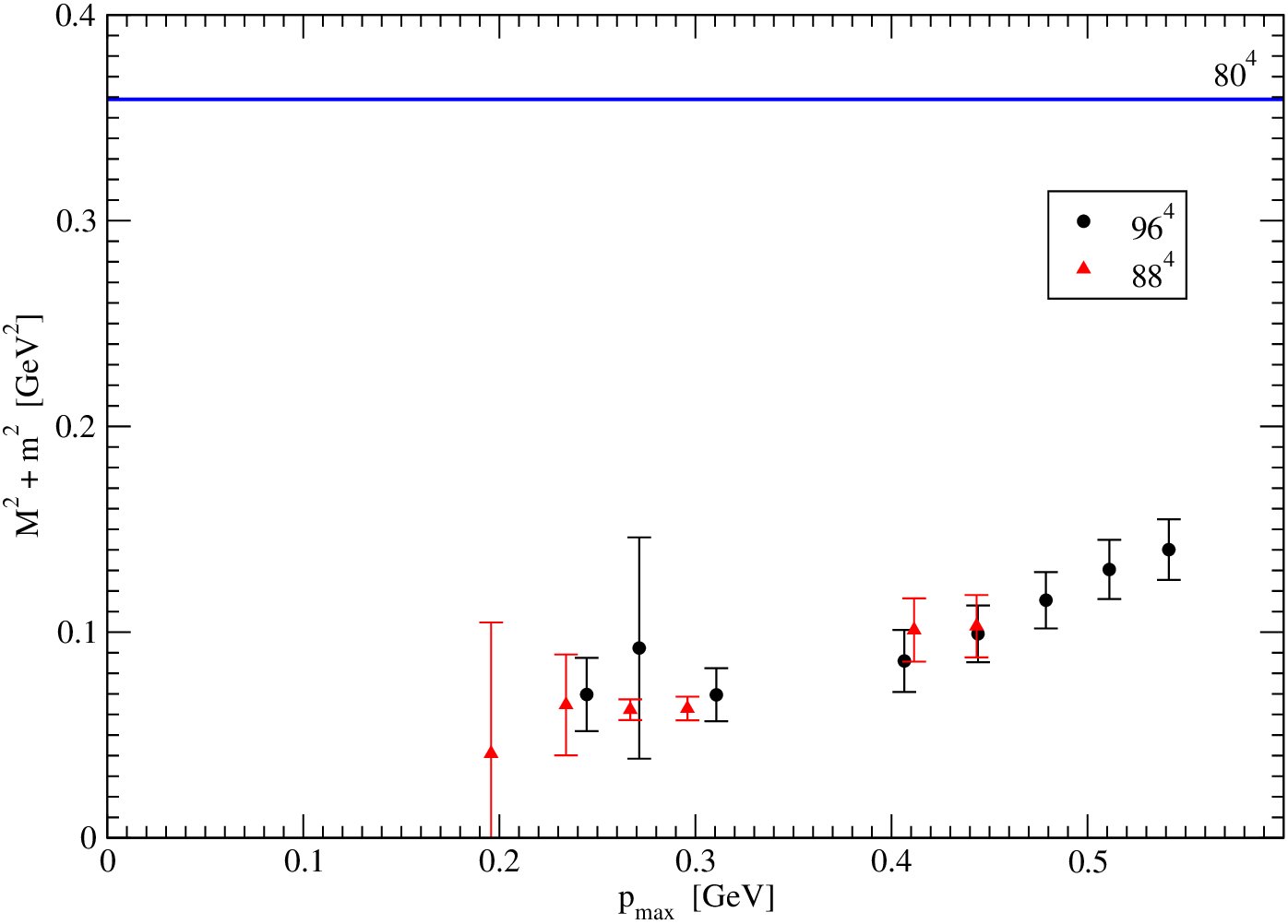} \\
   \vspace{0.9cm}
   \includegraphics[scale=0.3]{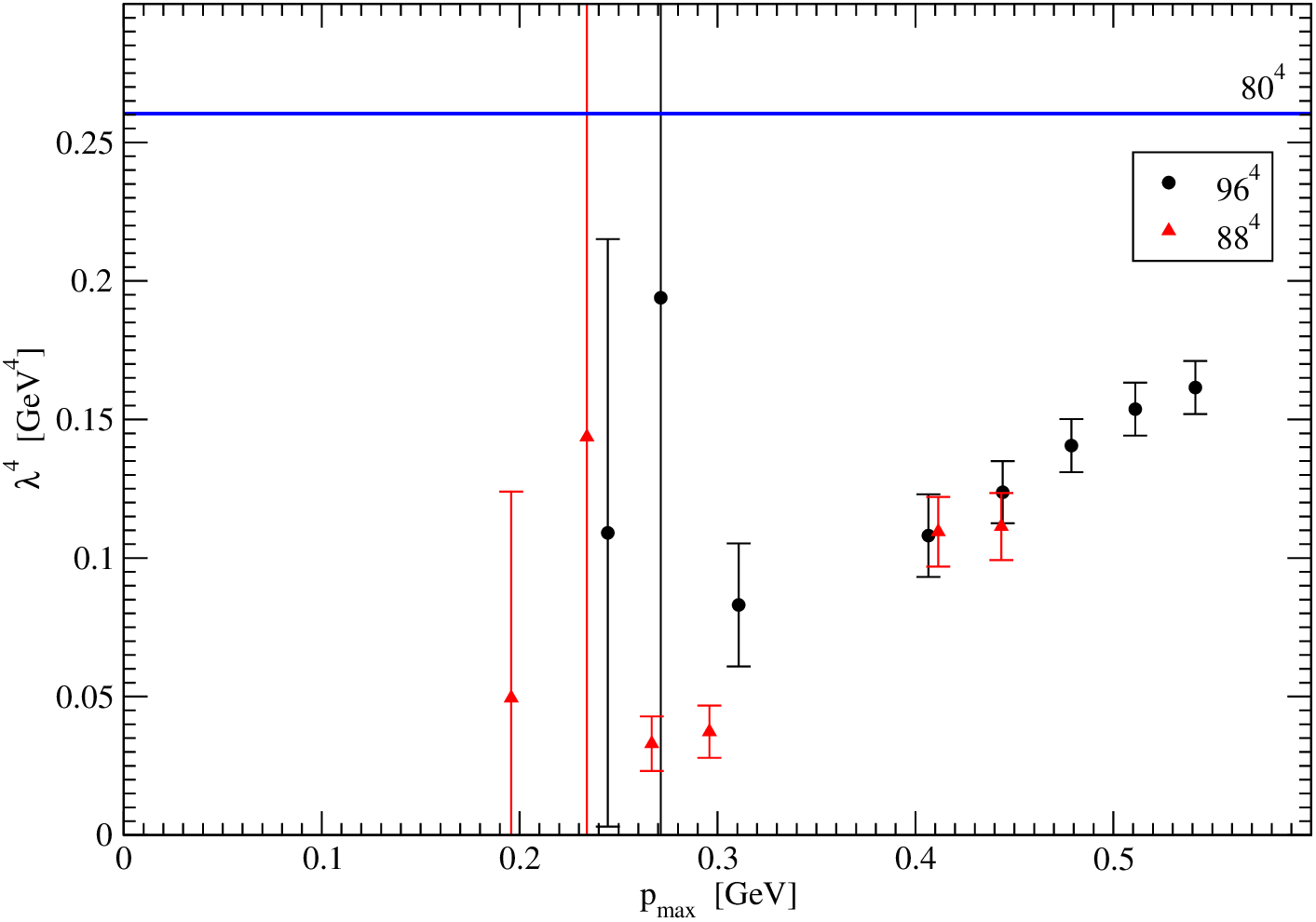}
      \caption{Results of fitting the $\beta = 5.7$ gluon propagator to (\ref{RZGtreelevel}) for the lattices $88^4$ and $96^4$. To guide the eye, we have included a full line showing the $\beta=5.7$, $80^4$ results, as reported in TABLE \ref{Fitsfinais}.}
   \label{Fitbeta57Large}
\end{figure}

In FIG.~\ref{Fitbeta57Large} we show the results of fitting the $88^4$ and $96^4$ propagators to equation (\ref{RZGtreelevel}) in the range
$[0, \, p_{max}]$ as a function of $p_{max}$. The plots only show the fitting parameters with a $\chi^2/d.o.f.$ smaller than three; we recall that, typically, one considers $\chi^2/d.o.f.$ below
two. As shown in the figure, the values are not stable against a change of $p_{max}$, in sharp contrast with the data shown in FIG.~\ref{Fitspmax64}. Moreover, looking at FIG.~\ref{Fitbeta57Large} we see that the values grow with increasing $p_{max}$ and seem to try to approach the typical numbers reported in TABLE \ref{Fitsfinais}. We call the reader's attention that in FIG.~\ref{Fitspmax64} a similar situation happens for the smallest fitting ranges. Indeed, only for $p_{max}$ larger than, let us say, $\sim
600$ MeV, the fitted parameter values start to become stable. In fact, also for all other lattice volumes we did consider in the main text, using smaller values of $p_{max}$ would give numerical values smaller than those reported in TABLE \ref{Fitsfinais}, while being unstable against variation around the chosen $p_{max}$.\\

The inability to fit the $88^4$ and $96^4$ data over wider momentum ranges can thus be explained by the infrared fluctuations. By performing infrared cuts, one could remove these fluctuations and fit the $88^4$ and $96^4$ propagators up to a $p_{max}$
well above 1 GeV. However, given that we want to discuss precisely the infrared region, we do not want to perform cuts at low momenta. Besides,
if one cuts the infrared data for those lattices, then one should also investigate its effect for all other lattices. Because of all that, we choose
not to include the $88^4$ and $96^4$ data in the analysis.

\begin{acknowledgments}
The work of O.~Oliveira is supported by FCT under project CERN/FP/83644/2008. D.~Dudal and N.~Vandersickel are supported by the Research-Foundation
Flanders (FWO Vlaanderen). We are grateful to S.~P.~Sorella for useful discussions.  We would also like to thank the Berlin, Moscow and Adelaide lattice groups for sending us their data and for allowing us to use it. We would in particular like to thank P.~J.~Silva for working out the gauge fixing and computing the
gluon propagator for the $32^4$ lattice at $\beta = 6.0$ and allowing us to use the data.
\end{acknowledgments}

\end{document}